# How Immersiveness Shapes the Link Between Anthropocentric Values and Resource Exploitation in Virtual Worlds

Quan-Hoang Vuong [1,2], Thi Mai Anh Tran [3], Ni Putu Wulan Purnama Sari [4], Fatemeh Kianfar [5], Viet-Phuong La [1], Minh-Hoang Nguyen [1,*]

[1] Centre for Interdisciplinary Social Research, Phenikaa University, Hanoi, Vietnam

[2] Professor, Korea University, Seoul, South Korea

[3] College of Forest Resources and Environmental Science, Michigan Technological University, Houghton, MI 49931 USA

[4] Faculty of Nursing, Widya Mandala Surabaya Catholic University, East Java, Indonesia

[5] Faculty of Humanities, University of Hormozgan, Bandar Abbas, Hormozgan, Iran

*Corresponding Email: hoang.nguyenminh@phenikaa-uni.edu.vn (Minh-Hoang Nguyen)

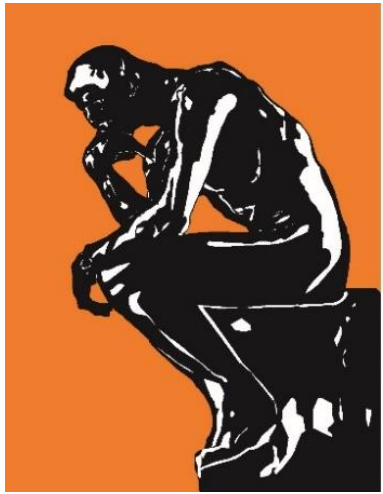

November 26, 2025

[*Original working draft v3 / Un-peer-reviewed*]

"[…] humans install new bodyguards that are great products of the digital technology era: robots. These humanoid figures—a deep learning-based Bird Image Identification System, using TensorFlow, and, made of endurable materials—are very sophisticated and stand in the fields in all their haughty superiority. More amazing is the fact that these robots are armed with various tools to combat incoming birds.

That seems to put an end to the birds' free ride on fresh vegetables and fruits…."

—In "Bogeyman"; *Wild Wise Weird* (2024)

Abstract

The Anthropocene is characterized by escalating ecological crises rooted not only in technological and economic systems but also in deeply ingrained anthropocentric worldviews that shape human–nature relationships. As digital environments increasingly mediate these interactions, video games provide novel contexts for examining the psychological mechanisms underlying environmental behaviors. This study investigates how anthropocentric values are associated with resource-exploiting behaviors in virtual ecosystems—specifically fishing, bug catching, and tree cutting—and how immersiveness moderates these relationships. Employing the Bayesian Mindsponge Framework (BMF) to analyze data from 640 Animal Crossing: New Horizons (ACNH) players across 29 countries, the study reveals complex links between anthropocentric worldviews and in-game behaviors. Fishing and tree-cutting frequencies are positively associated with anthropocentrism, whereas immersiveness weakens the association between tree cutting and anthropocentrism. Bug-catching frequency shows no direct effect but exhibits a growing negative association with anthropocentrism as immersiveness increases. These findings extend environmental psychology into virtual ecologies, illustrating how digital interactions both reflect and reshape environmental values. They highlight the potential of immersive gameplay to cultivate the Nature Quotient (NQ) and foster an eco-surplus culture through reflective, conservation-oriented engagement.

**Keywords:** Anthropocentrism; virtual ecology; immersion; environmental values; Nature Quotient; Granular Interaction Thinking Theory (GITT)

Introduction

The Anthropocene represents a global environmental crisis driven by human activity, characterized by climate change, rapid biodiversity loss, growing inequalities, and the erosion of Earth's capacity to absorb human-generated waste (Folke et al., 2021; Steffen et al., 2011). While the scale and urgency of these challenges are unprecedented, their roots extend beyond technological or economic factors to encompass deeply embedded psychological constructs that shape human-nature relationships (Dewi, Winoto, Achsani, & Suprehatin, 2025). Understanding these underlying psychological mechanisms is critical for developing effective interventions to address unsustainable behaviors that perpetuate environmental degradation (Hornsey, 2021; Lacroix & Gifford, 2017).

Environmental psychology research has consistently demonstrated that individuals' worldviews—their fundamental beliefs about reality and their place within it—significantly influence their environmental values, attitudes, and behaviors (Lacroix & Gifford, 2017). Among these worldview dimensions, anthropocentrism—the belief that value is human-centered and that all other beings are means to human ends—has been identified as particularly problematic for environmental sustainability (Fortuna, Wróblewski, & Gorbaniuk, 2023; Kopnina, Washington, Taylor, & J Piccolo, 2018; Nguyen, 2024; Tran et al., 2025). Studies have shown that stronger anthropocentric beliefs correlate with reduced environmental concern, greater endorsement of natural resource exploitation, and decreased engagement in pro-environmental behaviors (Aviste & Niemiec, 2023; Kaida & Kaida, 2016; Lou, Ito, & Li, 2025; Sockhill, Dean, Oh, & Fuller, 2022).

In today's digital era—where people, especially the younger generation, are exposed early to virtual environments and spend significant time immersed in video games—the formation of value systems is increasingly mediated by digital interactions. Within games, players continuously engage with information, make moral choices, and experience emotional connections that can shape their perceptions of nature and conservation (M.-T. Ho & Vuong, 2024). Such interactions can nurture empathy, appreciation for life, and pro-environmental attitudes, thereby reinforcing a "pro-conservation" value system (Vuong, Ho, Nguyen, et al., 2021). Therefore, understanding how people interact within games is no less important than studying their interactions in the physical world, as both realms influence the development of conservation-oriented mindsets and behaviors.

The digital age has given rise to immersive virtual worlds and video games, which function as sophisticated microcosms for observing human behavior (Fjællingsdal & Klöckner, 2017). With approximately 2.7 billion players worldwide (Fisher, Yoh, Kubo, & Rundle, 2021), gaming offers unprecedented opportunities to understand how individuals interact with simulated environments. These platforms offer unique methodological advantages over traditional laboratory experiments or surveys. They provide controlled yet ecologically dynamic settings and standardized scenarios that create engaging, naturalistic contexts (Blascovich et al., 2002; Loomis, Blascovich, & Beall, 1999). This structure allows researchers to observe actual, authentic behaviors unfolding over extended periods, yielding data that reflects genuine choices rather than self-reported intentions or responses to explicit experimental prompts (Yaremych & Persky, 2019). The COVID-19 pandemic amplified the significance of these virtual spaces. Animal Crossing: New Horizons (ACNH), released in March 2020, exemplifies this phenomenon, selling over 26 million units worldwide and becoming a cultural touchstone for millions during global lockdowns (Martinez et al., 2022). In ACNH, players cultivate islands, manage natural resources, interact with anthropomorphized animals, and make continuous decisions about environmental exploitation or conservation. The game's mechanics—including fishing, insect catching, tree harvesting, and habitat modification—create a simplified yet meaningful ecosystem in which players' environmental values can manifest in their virtual behaviors (Lewis, Trojovsky, & Jameson, 2021).

Initial research has begun to establish connections between virtual behaviors and real-world environmental attitudes. Vuong et al. (2021) and Ho et al. (2022) demonstrated, through analyses of ACNH gameplay data, that players' in-game exploitation behaviors correlated with their general environmental perceptions. Similarly, Tlili et al. (2024) found that players' environmental perceptions influenced their in-game behaviors toward virtual nature, though this relationship was not always consistent, as players sometimes acted against their stated beliefs to progress in the game. These findings suggest that virtual environments serve as revealing microcosms of human-nature relationships and as potential tools for conservation messaging and ecological learning (Coroller & Flinois, 2023; Fisher et al., 2021).

Despite parallel advances in environmental psychology and game studies, three critical gaps persist at their intersection. First, research lacks specificity in examining how anthropocentrism manifests in virtual contexts. While studies have documented relationships between general environmental attitudes and game behaviors (Tlili, Agyemang Adarkwah, et al., 2024; Vuong, Ho, Nguyen, et al., 2021), they typically aggregate anthropocentrism with broader environmental value orientations rather than isolating its unique predictive power. Anthropocentrism represents a distinct psychological construct—a fundamental worldview about human-nature hierarchies—yet its translation into quantifiable exploitative behaviors within virtual ecosystems remains unexamined. This specificity matters because anthropocentrism operates differently from other environmental attitudes: it reflects deep-seated beliefs about humanity's place in nature rather than surface-level preferences or concerns (Lou et al., 2025).

Second, the subjective experience of gameplay, particularly immersion, remains unexplored as a potential moderator between worldviews and virtual behaviors. This gap is theoretically significant because immersion could function in two opposing ways. On one hand, deeper immersion might amplify the expression of underlying worldviews. By fostering a state of psychological presence, players who feel genuinely "in" the game world may allow their fundamental beliefs about human-nature relationships to guide their choices more strongly. Research showing that embodying a tree in immersive virtual reality increases nature-relatedness supports this possibility (Spangenberger, Geiger, & Freytag, 2022).

On the other hand, immersion might attenuate the expression of worldview. This could occur if immersion shifts a player's focus from value-consistent behavior to game mechanics and progression goals. The finding that anti-anthropocentric players will harm virtual nature to advance in a game (Tlili, Agyemang Adarkwah, et al., 2024) suggests that immersion in the system may override personal values. This ambiguity is critical because immersion—characterized by absorption in the game and detachment from external reality (Michailidis, Balaguer-Ballester, & He, 2018) is a powerful psychological process. It is already known to enhance learning, intensify emotional responses, and strengthen the transfer of virtual experiences to real-world attitudes (Cheng, She, & Annetta, 2015; Y. Zhang & Song, 2022). Without investigating its moderating role, we cannot determine whether virtual environments ultimately amplify or suppress the behavioral manifestation of anthropocentric worldviews.

Third, existing analyses lack appropriate theoretical and analytical frameworks to model this complex psychological process. The relationship in question involves a stable trait-like worldview (anthropocentrism), a specific behavioral pattern (virtual exploitation), and a dynamic experiential state (immersiveness) that may conditionally moderate the worldview-behavior link. Traditional analytical approaches struggle to capture such multi-level conditional relationships, particularly when dealing with naturalistic behavioral data from commercial games.

To address this, our study employs the Granular Interaction Thinking Theory (GITT) for conceptual development and the Bayesian Mindsponge Framework (BMF) for statistical analysis of a dataset of 640 ACNH game players from 29 countries. The study aims to examine how anthropocentric worldviews manifest as virtual exploitation behaviors in ACNH and to determine whether this relationship is moderated by player immersiveness. Specifically, we address the following research questions:

- How are Animal Crossing: New Horizons (ACNH) game players' exploitation behaviors in the virtual world associated with their anthropocentrism mindset?
- Do game players' in-game immersiveness (i.e., perceived rich experience) moderate the association between the frequency of exploitation behaviors and their anthropocentrism?

Our study offers several key contributions. Theoretically, it pioneers a model of how abstract worldviews are associated with virtual actions, with immersiveness as a key moderating factor. Methodologically, we demonstrate the utility of the BMF for analyzing large-scale commercial game data, offering a new pathway for testing complex psychological theories outside the lab. Empirically, our analysis of a multinational dataset provides robust insights into human-nature dynamics, informing how game design can either reinforce or reshape real-world environmental values.

## Method

### *Theoretical foundation*

This study is grounded in Granular Interaction Thinking Theory (GITT), which explains how individuals' interactions with virtual ecosystems reflect and reshape their core values (Vuong & Nguyen, 2024a, 2024b). GITT conceptualizes the human mind as a dynamic information-processing system that seeks to maintain cognitive equilibrium. Within this framework, the informational entropy-based value paradigm posits that individuals continuously engage with their environment, absorbing and organizing informational inputs within the mind to reduce entropy through value formation (Vuong, La, & Nguyen, 2025a). These values subsequently serve as benchmarks for evaluating and filtering new information: inputs consistent with existing values are more readily accepted, whereas those that conflict are likely to be rejected to preserve cognitive stability (Vuong, La, & Nguyen, 2025b).

In this study, a player's interaction with the ACNH virtual ecosystem is considered to be associated with their core values concerning anthropocentrism. As players encounter environmental stimuli in the game world, their preexisting beliefs and values interact with new information, potentially updating them accordingly. When the in-game information strongly contradicts one's existing worldview, players may withdraw from play. However, if the information is perceived as compatible, neutral, or offset by compensatory benefits—such as social engagement or the avoidance of fear of missing out—players are likely to continue engaging. Through repeated in-game decisions involving resource extraction, habitat modification, and interspecies interaction, players acquire new information that can reinforce or reshape their value structures. Over time, these interactions may guide the mind toward a new cognitive equilibrium shaped by both preexisting beliefs and novel inputs absorbed from the virtual environment.

Within the GITT framework, anthropocentrism functions as a core value that both shapes and is shaped by in-game interactions. Players with stronger anthropocentric worldviews possess cognitive architectures that prioritize human-centered utility, perceiving virtual ecosystems primarily as resources for progress and accumulation. They process in-game entities—fish, insects, trees—through extractive filters that emphasize instrumental value. Conversely, players with weaker anthropocentric orientations may adopt alternative cognitive filters that attribute intrinsic worth to non-human entities, fostering more conservative or sustainable virtual behaviors.

Immersiveness acts as a critical moderating mechanism within this cognitive system. GITT defines immersiveness not merely as engagement intensity but as a state of heightened informational permeability, wherein the boundary between the player's cognitive system and the virtual environment becomes increasingly porous (Vuong, La, et al., 2025b). This psychological state influences the strength and direction of mind–environment interactions through two primary pathways.

In the amplification pathway, high immersiveness creates cognitive resonance that allows anthropocentric filters to operate with minimal interference from real-world constraints (Vuong, La, et al., 2025b). Deep immersion reduces cognitive dissonance, enabling the mind to internalize game-world information that aligns with preexisting beliefs. Extractive behaviors—catching rare fish, deforesting islands, accumulating materials—thus reinforce anthropocentric cognition through positive feedback loops, becoming legitimate expressions of human dominance and exploitation of nature.

Alternatively, in the attenuation pathway, immersion may redirect cognitive processing toward game-specific mechanics and objectives, temporarily overriding worldview-based decision-making. Highly immersed players may prioritize achievement, progression, or system mastery over value-consistent behavior. Under this mechanism, even those with weak anthropocentric worldviews may engage in exploitative actions, as the cognitive system adapts to game logic rather than real-world moral considerations.

The dynamics of this relationship reflect GITT's emphasis on context-dependent information processing. Virtual exploitation behaviors and anthropocentric tendencies are not mere outcomes of gameplay or static beliefs but arise from complex interactions among worldview values, transient psychological states, and environmental affordances. The virtual environment, in turn, generates informational feedback that can either reinforce or challenge existing cognitive filters.

Furthermore, GITT recognizes that virtual spaces create distinctive conditions for the expression of a worldview. Unlike the real world, where social norms, material limits, and awareness of consequences regulate behavior, virtual ecosystems offer relatively consequence-free environments where cognitive filters operate with minimal external constraints. This liberation allows anthropocentric attitudes and exploitative behaviors to manifest more directly, revealing the underlying informational logic of value formation in its "purest" form, moderated by immersion depth.

By integrating anthropocentric worldviews, exploitative gameplay behaviors, and immersive psychological states within the GITT framework, this study elucidates how virtual interactions can both mirror and transform human-nature relationships. This dynamic, multi-level perspective highlights virtual environments as both reflective spaces of preexisting worldviews and transformative arenas for reshaping environmental values. Understanding these cognitive mechanisms provides valuable insights for leveraging gaming platforms as tools for environmental education and for cultivating deeper, more sustainable relationships with nature.

*Materials and variables*

The study employed an observational, cross-sectional design using a dataset of 640 Animal Crossing: New Horizons (ACNH) players from 29 countries worldwide. This peer-reviewed dataset has been openly deposited in *Data Intelligence* (Vuong, Ho, La, et al., 2021). Data were collected between May 15 and May 30, 2020, via an online survey administered via Google Forms and distributed within ACNH player communities on Discord, Reddit, and Facebook. Google Forms was chosen for its accessibility, confidentiality, and ease of dissemination through shareable links. Prior to distribution, administrators or moderators of each community were contacted for approval, and the study's objectives, contents, and compliance with community rules were clearly explained. The survey post—containing a detailed description of the research and the questionnaire link—was shared only after formal permission was granted. Participants were required to read and agree to an informed consent statement before participating. As a token of appreciation, the first 100 respondents received a US$5 Amazon gift card, while the next 200 received a US$2 card.

Before the main survey, a pilot test was conducted with 15 students from Japan, Singapore, the United States, and Vietnam to ensure the clarity and reliability of the questionnaire. Feedback from this pilot informed minor revisions for better precision. To minimize missing data, all questions were set as mandatory, preventing submission of incomplete responses. Contact information was provided for real-time support, and all participant inquiries were addressed promptly. Upon completion, an official announcement marked the end of data collection across all communities. Responses were exported from Google Forms in both Microsoft Excel (.xls) and comma-separated values (.csv) formats. Data cleaning involved clarifying ambiguous responses, coding all variables, and validating results through visualization and statistical checks.

Among the 640 respondents, 64.38% (n = 412) were female. Participants were primarily from the United States and Canada (55%), followed by Asia (28.13%) and the European Union (14.38%), with the remaining 2.5% representing other regions. Ethnic distribution mirrored this pattern, with White (54.22%) and Asian (31.25%) players comprising the majority. Over half of the respondents were undergraduate students (52.5%), and most were single and never married (61.09%). The majority were young adults aged 18–30 years (72.8%), with ages ranging from 11 to 55. Additionally, 63% (n = 403) reported having stable employment, and only 13.44% indicated not owning a pet or garden.

In this study, five variables were extracted from the original dataset, comprising one outcome variable and four predictors. The outcome variable, *Anthropocentrism*, was derived from variable *C12*—item 12 in the revised New Ecological Paradigm Scale (NEPS) developed by Dunlap et al. (2000). This item is part of the NEPS subscale assessing anti-anthropocentrism, which includes three related statements. However, the internal consistency test for this subscale yielded a relatively low Cronbach's α (0.56), likely reflecting both the conceptual complexity of anthropocentrism and the cultural diversity of the international sample. Owing to this low internal reliability, the present study focused solely on item 12, which gauges respondents' agreement with the statement, "Humans were meant to rule the rest of nature" (White, 1967; Fleck, 2001). Responses were recorded on a five-point Likert scale ranging from 1 ('strongly disagree') to 5 ('strongly agree').

To investigate players' resource-exploiting behaviors in ACNH, participants were asked to report the frequency of three main in-game resource exploitation activities—catching bugs, fishing, and cutting down trees—on a four-point Likert scale ranging from 1 (never) to 4 (often). These behavioral variables were generated from variables *E1*, *E2*, and *E17* in the original dataset, respectively. Additionally, immersiveness was measured using variable *F30*, which represents the player's feeling of having a rich experience while playing the game. Although this item belongs to the Sensory and Imaginative Immersion subscale of the Game Experience Questionnaire—which includes six items addressing multiple aspects such as story, aesthetics, imagination, exploration, and impressiveness—only this specific variable was selected to isolate the sense of immersive experience during gameplay for the purpose of this study.

### Table 1. Variable Description

| Variable Name (Name in the original dataset) | Description | Data Type | Measurement |
|---|---|---|---|
| *Anthropocentrism* (*C12*) | Self-reported agreement with the following statement: "Humans were meant to rule over the rest of nature." | Numerical | 5-point Likert scale (1 = Strongly Disagree, 5 = Strongly Agree). |
| *CatchBug* (*E1*) | Self-reported frequency of catching insects. | Numerical | 4-point Likert scale (1 = Never, 4 = Often). |
| *Fishing* (E2) | Self-reported frequency of fishing activities. | Numerical | 4-point Likert scale (1 = Never, 4 = Often). |
| *CutDownTree* (E17) | Self-reported frequency of cutting down trees. | Numerical | 4-point Likert scale (1 = Never, 4 = Often). |
| *RichExperience* (F30) | Self-reported feeling of rich experience while playing game. | Numerical | 5-point Likert scale (1 = Not at all, 5 = Extremely). |

### Statistical model

To examine the research question, we developed the following Model to analyze the relationships between game players' exploitation behaviors in the virtual world and their anthropocentrism mindset, as well as whether game players' in-game immersiveness (i.e., perceived rich experience) moderates the association between the frequency of exploitation behaviors and their anthropocentrism.

$$Anthropocentrism \sim normal(\mu, \sigma) \qquad (1.1)$$

$$\mu_i = \beta_0 + \beta_1 * CatchBug_i + \beta_2 * Fishing_i + \beta_3 * CutDownTree_i + \beta_4 * RichExperience_i + \beta_5 * CatchBug_i * RichExperience_i + \beta_6 * Fishing_i * RichExperience_i + \beta_7 * CutDownTree_i * RichExperience_i \qquad (1.2)$$

$$\beta \sim normal(M, S) \qquad (1.3)$$

The probability around $\mu$ is determined by the form of the normal distribution, whose width is specified by the standard deviation $\sigma$. $\mu_i$ represents game player $i$'s level of antrhopocentrism. The variables $CatchBug_i$, $Fishing_i$, $CutDownTree_i$, and $RichExperience_i$ correspond to game player $i$'s frequency of catching bug, fishing, cutting down trees, and feeling of immersiveness while playing game, respectively. Meanwhile, coefficients $\beta_5$-$\beta_7$ indicate the non-additive effects of $CatchBug_i * RichExperience_i$, $Fishing_i * RichExperience_i$, and $CutDownTree_i * RichExperience_i$ on $Anthropocentrism$. The model has nine parameters, including an intercept of $\beta_0$, coefficients of $\beta_1$-$\beta_7$, and the standard deviation of the "noise", $\sigma$ (i.e., uncertainty or variability in the observed data that is not explained by the model). The coefficient values follow a normal distribution with a mean denoted by $M$ and a standard deviation denoted by $S$.

### Data analysis and validation

The present study employed the Bayesian Mindsponge Framework (BMF) analytics for several reasons. First, BMF synthesizes the reasoning strength of GITT with the computational inference power of Bayesian analysis, making it particularly suitable for analyzing complex and uncertain psychological processes (Nguyen, La, Le, & Vuong, 2022; Vuong, Nguyen, & La, 2022). Second, Bayesian inference treats both known and unknown parameters as probabilistic entities (Csilléry, Blum, Gaggiotti, & François, 2010; Gill, 2014), thereby supporting the estimation of parsimonious models. Through the Markov Chain Monte Carlo (MCMC) technique, Bayesian analysis also efficiently handles complex models, including multilevel and nonlinear structures (Dunson, 2001). Third, compared with the frequentist approach, Bayesian inference provides several advantages—most notably, the use of credible intervals that enable more nuanced and continuous interpretation of uncertainty, rather than relying solely on binary significance testing based on p-values (Halsey, Curran-Everett, Vowler, & Drummond, 2015; Wagenmakers et al., 2018).

During model construction, prior selection is a crucial step in Bayesian analysis. Given the exploratory nature of this study, uninformative (flat) priors were adopted to minimize prior influence on model estimation (Diaconis & Ylvisaker, 1985). To assess the robustness of posterior distributions, a prior-tweaking technique was employed by re-running the analysis with a prior that represents disbelief in the associations (a normal distribution with mean = 0 and SD = 0.5). If posterior estimates remain almost identical to those obtained with uninformative priors, the results can be considered insensitive to prior assumptions. Furthermore, incorporating priors helps resolve multicollinearity. As Leamer (1973) notes, in Bayesian inference, multicollinearity reflects a "weak data problem" characterized by large standard errors and the overlap between prior and posterior distributions within certain subspaces. When prior information is sufficiently informative, this issue can be mitigated effectively. Empirical evidence supports this claim, showing that Bayesian models with informative priors outperform Ridge regression in addressing multicollinearity (Adepoju & Ojo, 2018; Jaya, Tantular, & Andriyana, 2019).

After model estimation, the Pareto-smoothed importance sampling leave-one-out (PSIS-LOO) diagnostic was employed to evaluate model goodness-of-fit (Vehtari & Gabry, 2019; Vehtari, Gelman, & Gabry, 2017). The LOO statistic is calculated as:

$$LOO = -2LPPD_{loo} = -2\sum_{i=1}^{n} \log \int p(y_i|\theta)p_{post(-i)}(\theta)d\theta$$

Where $p_{post(-i)}(\theta)$ denotes the posterior distribution estimated using all data points except observation $i$. In the PSIS method, *k*-Pareto values are used to identify observations exerting disproportionate influence on the LOO estimate. Cases with *k*-values greater than 0.7 are typically considered influential and may bias cross-validation accuracy, whereas values below 0.5 generally indicate a well-fitting model.

Upon confirming model fit, convergence diagnostics were performed prior to result interpretation. Convergence of the Markov chains was assessed using both statistical and visual approaches. Statistically, two key indicators were examined: the effective sample size (*n_eff*) and the Gelman–Rubin shrink factor (*Rhat*) (Brooks & Gelman, 1998). The *n_eff* value represents the number of effectively independent samples, free of autocorrelation, during the stochastic simulation, while *Rhat* reflects the potential scale reduction factor across chains. Typically, *n_eff* > 1000 suggests sufficient convergence and sampling adequacy (McElreath,

2018), and models are considered convergent when *Rhat* ≈ 1; values above 1.1 indicate non-convergence. Visually, convergence was assessed through trace plots, which showed that the chains mixed well and exhibited stable posterior distributions.

All Bayesian analyses were conducted in R using the open-access package bayesvl version 1.0.0 (La & Vuong, 2019; Vuong & La, 2025), which offers extensive visualization capabilities and user-friendly operationalization. To promote transparency and reproducibility, all data and analytical scripts have been archived on Zenodo: https://zenodo.org/records/17472912

## Results

The PSIS-LOO diagnostics indicated that all *k*-values were below the threshold of 0.5, confirming that the constructed model exhibited a good fit with the dataset (see Figure 1). Therefore, the estimated results can be reliably interpreted.

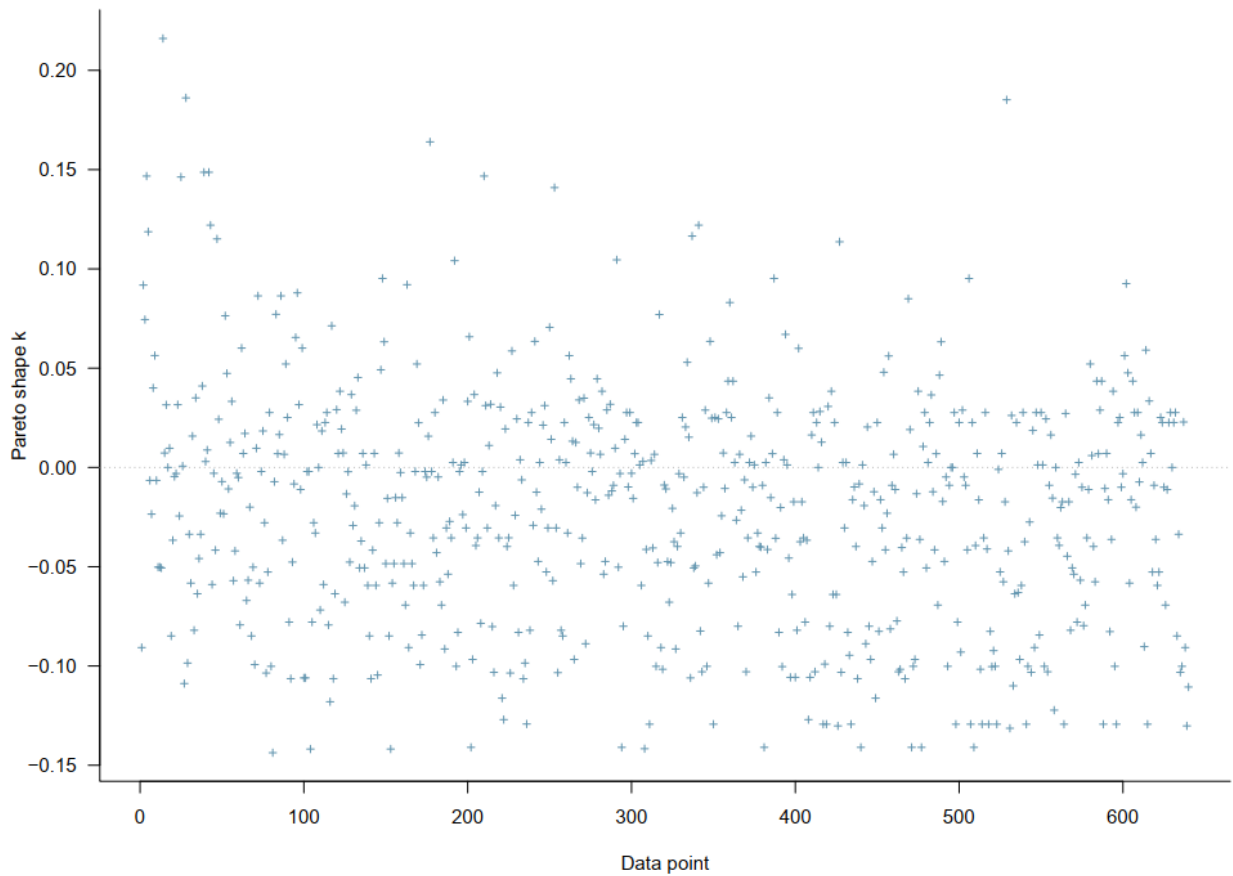

**Figure 1.** PSIS-LOO diagnostic plot estimated using uninformative priors

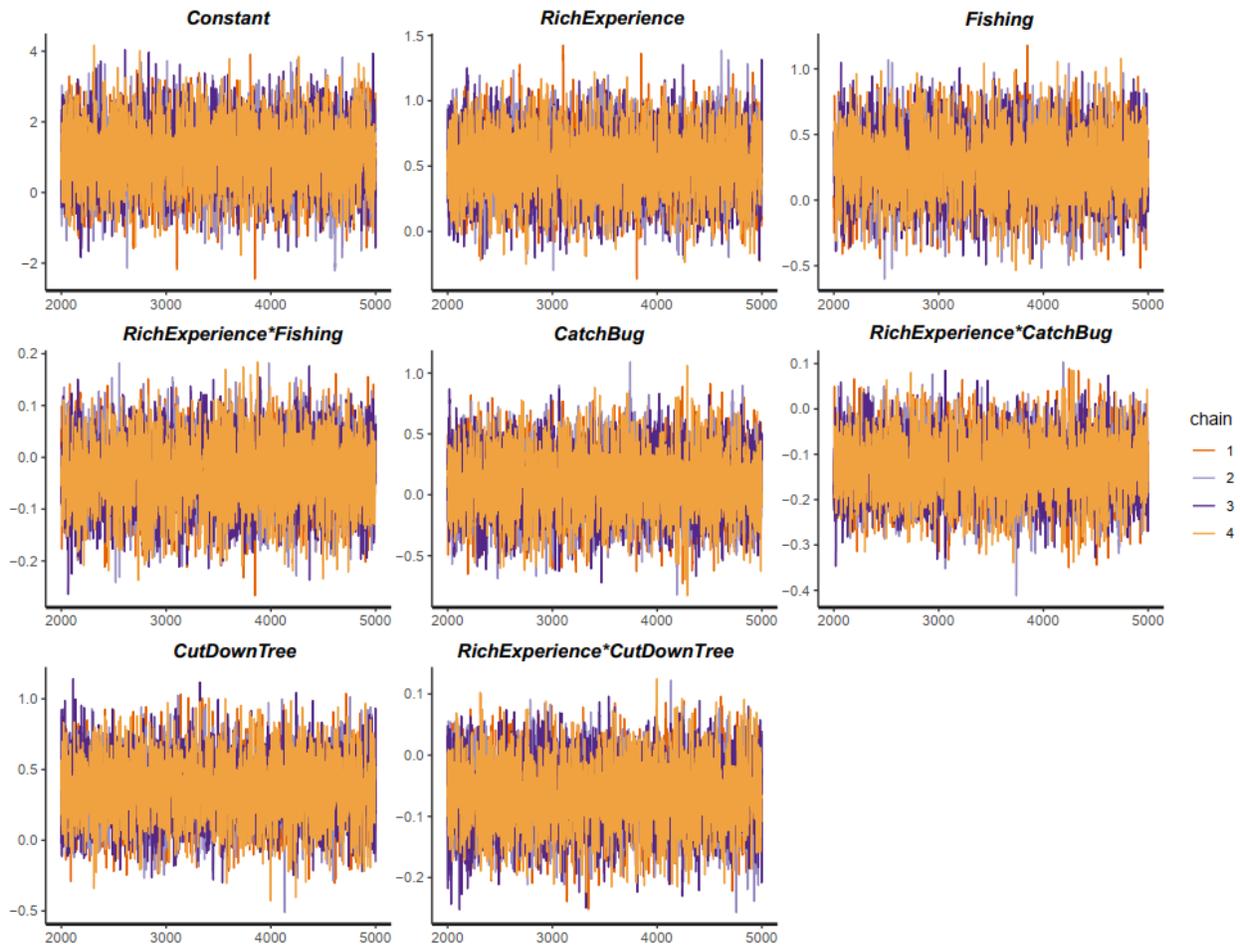

**Figure 2.** Trace plots estimated using uninformative priors

As presented in Table 2, all estimated posterior parameters of the analytical model achieved satisfactory convergence diagnostics, with *n_eff* values exceeding 1,000 and *Rhat* values equal to 1. These results indicate that the Markov chains achieved proper convergence, providing sufficient effective sampling for reliable inference. Visual diagnostics in Figure 2 further corroborate this finding, as the trace plots display well-mixed chains fluctuating stably around their central equilibria.

**Table 2.** Estimated posteriors

| Parameters | Uninformative priors | | | | Informative priors (Reflecting disbelief) | | | |
|---|---|---|---|---|---|---|---|---|
| | M | S | n_eff | Rhat | M | S | n_eff | Rhat |
| Constant | 1.05 | 0.89 | 4621 | 1 | 1.57 | 0.77 | 4152 | 1 |
| RichExperience | 0.51 | 0.24 | 4824 | 1 | 0.37 | 0.20 | 4752 | 1 |
| Fishing | 0.27 | 0.24 | 4922 | 1 | 0.20 | 0.21 | 5362 | 1 |
| RichExperience*Fishing | -0.03 | 0.06 | 4713 | 1 | -0.01 | 0.05 | 5145 | 1 |
| CatchBug | 0.10 | 0.25 | 4321 | 1 | 0.07 | 0.21 | 5539 | 1 |
| RichExperience*CatchBug | -0.13 | 0.06 | 4736 | 1 | -0.12 | 0.06 | 5822 | 1 |
| CutDownTree | 0.36 | 0.21 | 5230 | 1 | 0.27 | 0.19 | 5316 | 1 |
| RichExperience*CutDownTree | -0.06 | 0.05 | 5317 | 1 | -0.04 | 0.05 | 5826 | 1 |

\* Note: M = Mean, S = Standard deviation, n_eff = effective sample size, and Rhat = Gelman shrink factor

The estimated results using uninformative priors show that feeling immersiveness in the game is positively associated with the anthropocentric level ($M_{RichExperience}$ = 0.51 and $S_{RichExperience}$ = 0.24). The frequency of fishing and cutting down trees is also positively associated with the anthropocentric level ($M_{Fishing}$ = 0.27 and $S_{Fishing}$ = 0.24; $M_{CutDownTree}$ = 0.36 and $S_{CutDownTree}$ = 0.21). Meanwhile, the frequency of catching bugs has an ambiguous relationship with the anthropocentric level ($M_{CatchBug}$ = 0.10 and $S_{CatchBug}$ = 0.25). As for the non-additive effects, both $RichExperience * CatchBug$ and $RichExperience * CutDownTree$ have negative association with $Anthropocentrism$ ($M_{RichExperience*CatchBug}$ = -0.13 and $S_{RichExperience*CatchBug}$ = 0.06; $M_{RichExperience*CutDownTree}$ = -0.06 and $S_{RichExperience*CutDownTree}$ = 0.05), whereas the effect of $RichExperience * Fishing$ is ambiguous ($M_{RichExperience*Fishing}$ = -0.03 and $S_{RichExperience*Fishing}$ = 0.06). The estimated results using informative priors reflecting our disbelief in the associations reveal patterns similar to those using uninformative priors, suggesting the findings are robust to changes in priors.

The posterior distributions of all coefficients are shown in Figure 3 along with their Highest Posterior Density Interval (HPDI) at 90%, indicating that 90% of the probability that the true parameter value lies within this range, given the data and the model. As can be seen, the HPDIs of $RichExperience$ and $CutDownTree$ are located entirely on the positive side of the x-axis, suggesting that their positive associations with $Anthropocentrism$ are highly reliable. Although not all 90% HPDI of $Fishing$ lies on the positive side, the portion situated on the negative side is negligible, and its mean value is larger than its standard deviation, so the positive association of $Fishing$ can be deemed highly reliable. The 90% HPDI of $RichExperience * CatchBug$ is located entirely on the negative, implying its highly reliable association. Meanwhile, although a portion of $RichExperience * CutDownTree$'s HPDI is still located on the positive side, that portion is minimal, and its mean value is larger than the standard deviation, so its negative association is considered highly reliable.

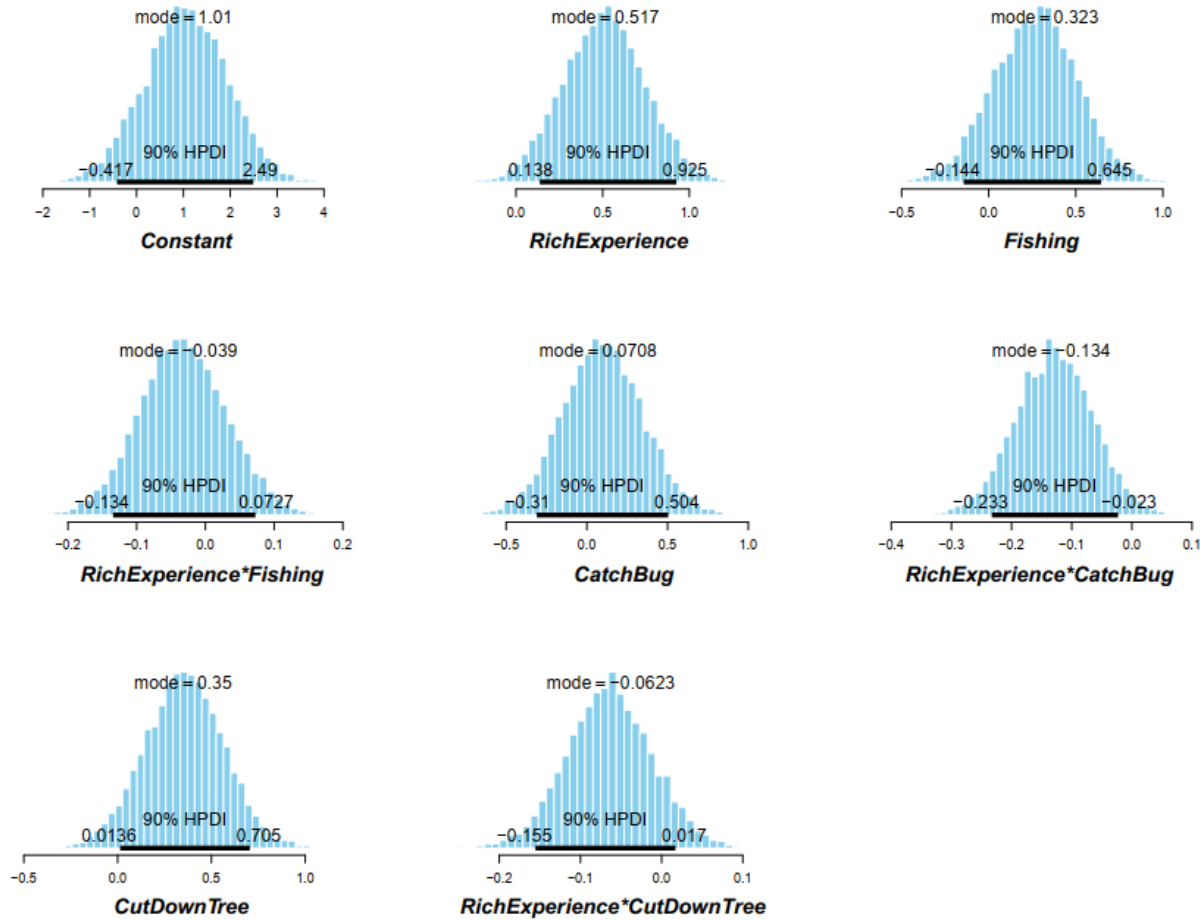

**Figure 3.** Estimated posterior distributions using uninformative priors

Employing Equation (1.2) and the estimated parameters' mean values from Table 2, we calculated the level of anthropocentrism among game players based on the frequency of their in-game activities and level of immersiveness. The estimated anthropocentrism levels corresponding to the frequencies of fishing, tree cutting, and bug catching are illustrated in Figures 4-a, 4-b, and 4-c, respectively.

Figure 4-a shows that players' anthropocentrism tends to increase with greater fishing frequency. Although the effects of different levels of immersiveness converge at higher frequencies of fishing (i.e., "sometimes" and "often"), the general tendency remains that a stronger feeling of immersiveness is associated with higher anthropocentrism.

Figure 4-b indicates that the frequency of cutting down trees is generally positively associated with anthropocentrism, but the influence of immersiveness varies across different frequency levels. Specifically, at low frequencies of tree cutting (i.e., "never" and "seldom"), players who feel more immersed are more likely to exhibit anthropocentric tendencies. Conversely, at high frequencies (i.e., "often"), those with greater immersiveness are less likely to be anthropocentric.

Figure 4-c clearly demonstrates that the relationship between bug-catching frequency and anthropocentrism is conditional on immersiveness. Among players who frequently catch bugs (i.e., "sometimes" and "often"), anthropocentrism decreases as immersiveness increases. In contrast, for players with low bug-catching frequency (i.e., "never" and "seldom"), anthropocentrism increases with higher immersiveness.

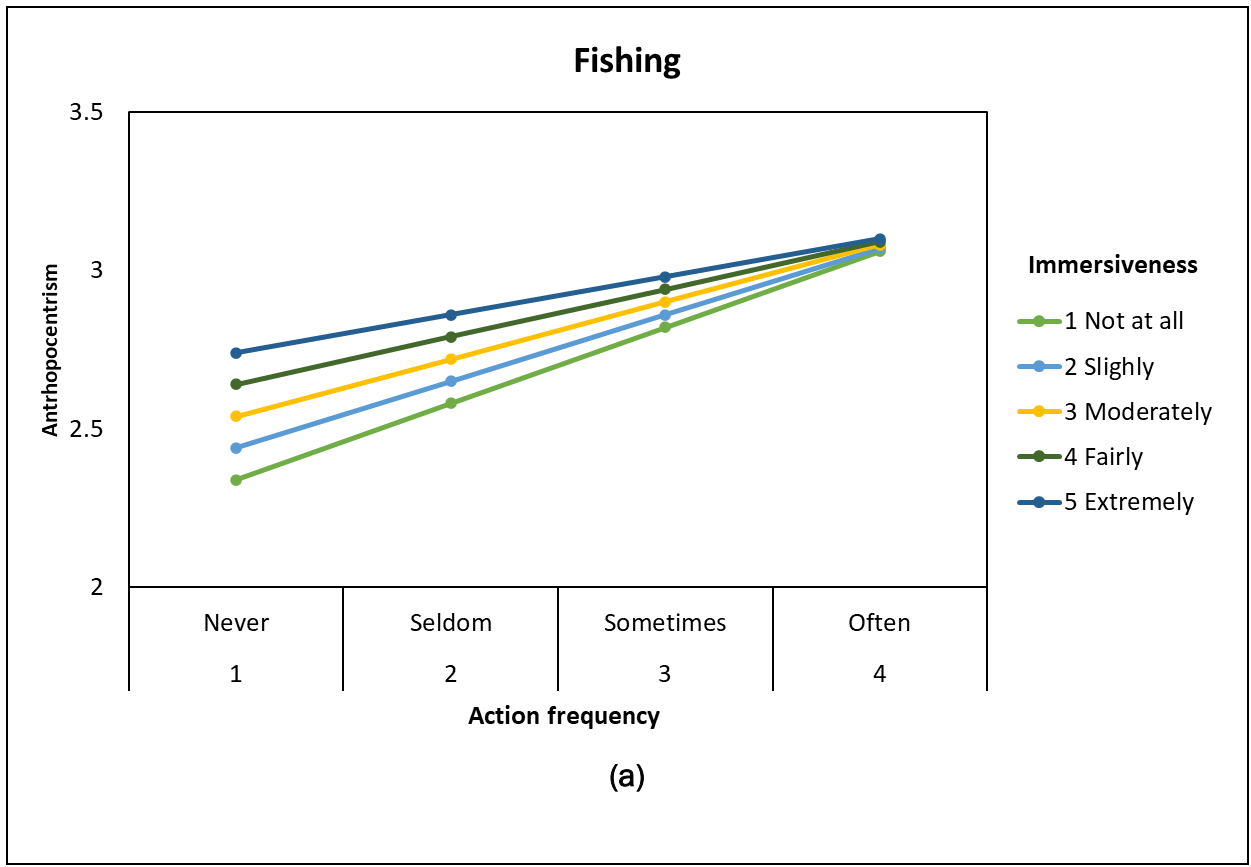

(a)

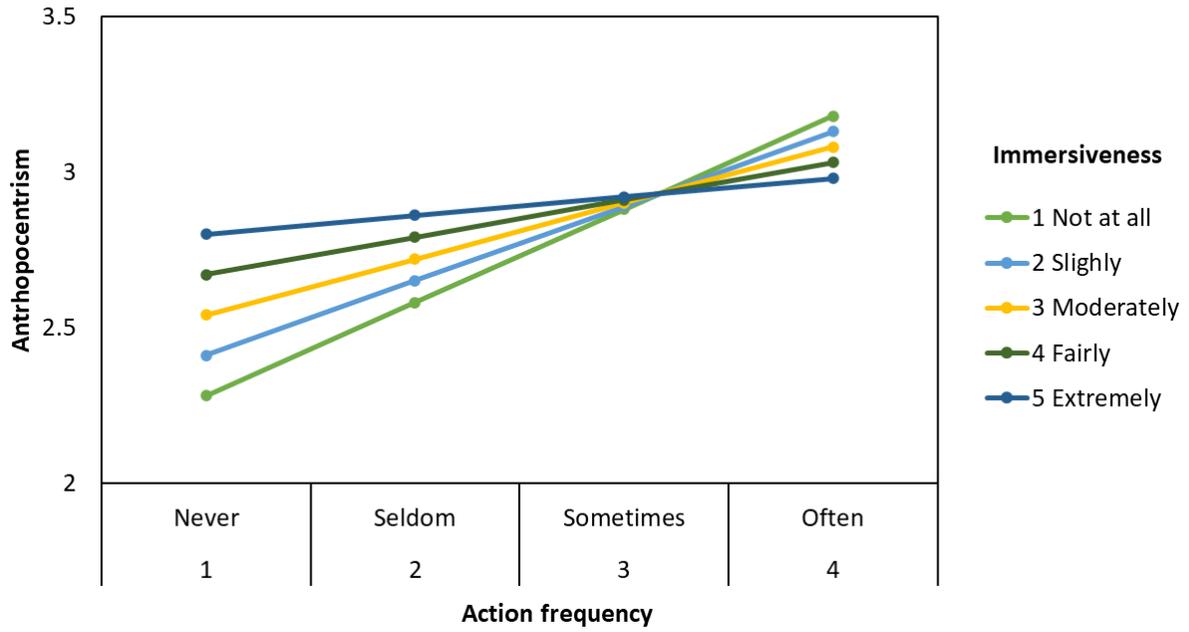

(b)

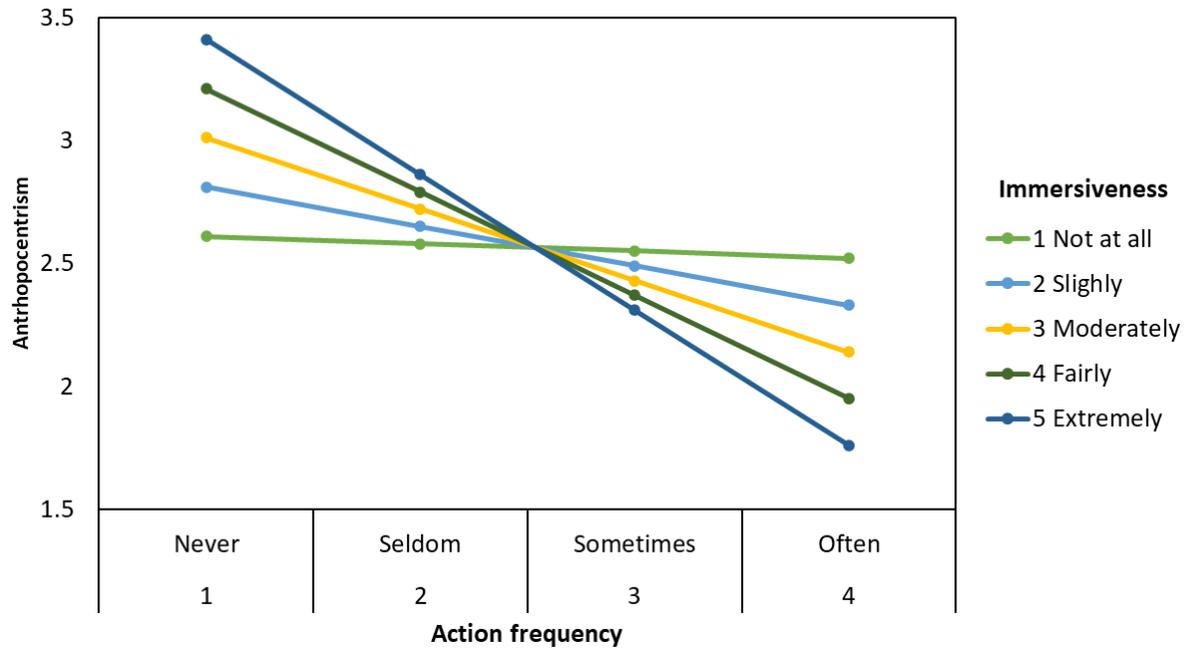

(c)

**Figure 4:** Estimated anthropocentrism level of game players based on their immersiveness while playing and the frequency of a) fishing, b) cutting down trees, and c) catching bugs.

Discussion

This study investigates how players' exploitative in-game actions—such as fishing, bug catching, and tree cutting—relate to anthropocentric values, and whether players' sense of immersiveness moderates these relationships. Using the Bayesian Mindsponge Framework (BMF) and a multinational dataset of 640 players from 29 countries, the study yields several notable findings.

*Virtual Fishing and Anthropocentrism*

The results reveal that the frequency of fishing is positively associated with anthropocentrism. In ACNH, fishing is designed to be a rewarding, consequence-free activity. As a result, frequent engagement in this virtual behavior can facilitate the absorption of information that subtly fosters or reinforces anthropocentric value orientations within players' minds. Cognitively, the act of fishing implicitly frames aquatic life as a resource to be collected, sold, or displayed—an instrumental interaction that reflects a human-centered utility. The commodification mechanism also mirrors real-world extractive behavior, reinforcing a utilitarian worldview in which fish are valued primarily for human use or entertainment (Udoudom, 2021).

Fish in ACNH are rendered as non-sentient collectibles rather than sentient beings who, in reality, are capable of conscious experience (Falk, 2024). They exhibit no fear, struggle, or autonomy; their existence is functionally reduced to player interaction. Such design choices reinforce an anthropocentric experience in which nature is conceived as a system of rewards for human satisfaction rather than as a network of entities possessing intrinsic value or agency. By omitting any signs of resistance or emotion, the game does not demonstrate the fish's perspective, dulling moral reflection. Consequently, players rarely question the act of fishing, perceiving it as an innocuous pastime rather than a simulated act of taking life.

Moreover, the game's reward structure incentivizes fishing through multiple extrinsic reinforcements—bells (in-game currency), museum prestige, and Critterpedia completion (the encyclopedia of all catchable species). These reward systems encourage repetitive, goal-driven behaviors that prioritize human benefits over ecological balance—an essential feature of real ecosystems, though absent in ACNH (Verma, 2017). Unlike real-world environments, overfishing in the game produces no depletion or biodiversity loss, removing ecological consequences that would normally evoke reflection. This lack of negative feedback may desensitize players to environmental impact, normalizing anthropocentric dominance through virtual behavioral conditioning.

As habitualization occurs, repeated fishing may become a normalized routine that consolidates values through reinforcement learning. Over time, players may internalize an anthropocentric view of nature that extends beyond the game environment. Prior studies indicate that virtual behaviors can shape real-world attitudes and actions (Burrows & Blanton, 2016). Furthermore, ACNH players often act contrary to their real-world environmental values to achieve progress within the game. For instance, environmentally conscious players may still harm virtual ecosystems to advance levels or collect rewards (Tlili, Adarkwah, Salha, & Huang, 2024).

*Virtual Tree-cutting and Anthropocentrism*

The study's findings indicate that the frequency of tree cutting is positively associated with anthropocentrism, but the feeling of immersiveness negatively moderates this relationship. In interpretive terms, players who frequently cut down trees and feel highly immersed in the game tend to exhibit higher anthropocentric values overall; however, at higher frequencies of tree cutting (i.e., 'often'), greater immersiveness is associated with lower anthropocentrism.

In ACNH, tree cutting is designed as a rewarding activity. The game frames trees primarily through their instrumental value—as sources of wood, bells, or construction space—reflecting a resource-extraction mentality with limited opportunities for alternative, non-utilitarian interactions. Players are encouraged to shape the island according to their aesthetic preferences, transforming tree removal into a form of personal expression rather than ecological care (Karlsson & Ryberg, 2024). This customization-over-conservation paradigm enhances player satisfaction while marginalizing ecological integrity values.

Moreover, the absence of ecological feedback in ACNH's mechanics makes deforestation consequence-free. Tree removal does not visibly disrupt ecosystems, wildlife, or biodiversity, detaching the virtual environment from real-world ecological interdependence. Trees also do not regenerate naturally once cut down; they must be replanted manually. This mechanic reinforces an illusion of control and reversibility, suggesting that environmental manipulation can be undone instantly and without cost—a stark contrast to real ecosystems, where regeneration is slow, uncertain, and context-dependent. Such design elements may desensitize players to environmental degradation and normalize human dominance over nature, reinforcing the perception that the environment is malleable and subordinate (Guanio-Uluru, 2021).

Frequent engagement in tree-cutting activities can thus enable the absorption and processing of information that can reduce informational entropy toward anthropocentric value orientations, subtly reinforcing human-centered worldviews within players' minds.

However, the negative moderation effect of immersiveness suggests that highly immersed players who frequently cut down trees may begin to recognize, albeit indirectly, the negative consequences of deforestation. This recognition may stem from the game's island evaluation system, which rates islands based on their level of development and scenery. Achieving higher ratings—and unlocking new milestones—requires maintaining a balanced number of trees, flowers, and shrubs to enhance scenery value. For more immersive players who care about the island's progress and aesthetics, this system functions as a subtle ecological feedback mechanism: cutting down too many trees reduces the island's rating, encouraging restoration. Thus, among players who often engage in tree cutting, greater immersiveness is associated with lower anthropocentrism, as these players perceive the costs of environmental imbalance within the game.

Conversely, this moderating effect does not appear among players with low tree-cutting frequency. One possible explanation lies in the upper threshold of the rating system, which penalizes an excessive number of trees. When the island exceeds roughly 220 mature trees, the evaluation system describes it as suffering from an "overabundance of trees" and warns that "people getting lost in the woods and fearing for the worst" (Animal Crossing Fandom, n.d.). In such cases, players must remove trees to restore balance and achieve the optimal rating. Thus, both excessive deforestation and overforestation are discouraged through feedback loops, indirectly shaping how immersive players internalize the ecological values embedded within ACNH's design.

*Virtual Bug-Catching and Anthropocentrism*

The study found that bug-catching frequency is not directly associated with anthropocentrism; rather, its effects emerge under different conditions of immersiveness. Specifically, for players who report no sense of immersiveness, the frequency of bug catching exerts a neutral effect on anthropocentric values. However, as immersiveness increases, the differences in anthropocentrism across varying frequencies of bug catching become more pronounced. Among highly immersive players (i.e., those reporting "fairly" or "extremely" immersive experiences), individuals who seldom or never catch bugs exhibit the highest levels of anthropocentrism, whereas those who catch bugs frequently ("sometimes" or "often") display the lowest. Among moderately and slightly immersive players, this negative association persists but with a smaller magnitude. These findings suggest that a certain in-game design or experiential mechanism may, under conditions of heightened immersiveness, reduce the extent to which players perceive themselves as the center of the virtual ecosystem.

Unlike fishing or tree cutting, bug catching in ACNH can foster less anthropocentric orientations because the activity requires close attention to ecological details. The game features over 80 species of bugs, each with unique habitats, behaviors, seasonal cycles, and spawning conditions that mimic ecological processes—for example, crickets, bagworms, and tarantulas spawn on the ground or in trees, while butterflies, mantises, and ladybugs depend on blooming flowers. For specific species to appear, players must create or preserve appropriate environmental conditions. Through such micro-ecological interactions, players engage in a form of ecological stewardship that can promote ecocentric rather than anthropocentric value orientations.

Bug catching also demands active observation, patience, and stealth. Players must carefully monitor the environment, move slowly, and time their actions precisely to succeed. This cultivates attentiveness to nonhuman life and fosters a sense of coexistence rather than dominance. Moreover, bug catching is one of the few activities in ACNH that introduces consequences for exploiting nature. Certain species, such as tarantulas and scorpions, are designed to be aggressive. When approached too closely or handled recklessly, they can chase and sting the player, causing the character to faint and reappear at home or the island's entrance. Such events provide a subtle but meaningful form of ecological feedback, reminding players of their limited control over nature and the agency of other species (Raihani & Bshary, 2019).

However, these ecological feedback mechanisms are relatively rare and subtle, as the game's overall design emphasizes calm, nonviolent interaction with nature. Consequently, the absorption and internalization of ecocentric information occur primarily among players who are deeply immersed in the game world. This pattern aligns with GITT's reasoning, which conceptualizes immersiveness as a state of heightened informational permeability—where the boundary between the player's cognitive system and the virtual environment becomes increasingly porous (Vuong, La, et al., 2025b). Indeed, immersive experiences heighten users' sense of presence and emotional engagement (Shin, 2019) and often involve sensory isolation from external stimuli, enhancing focus and reflection (C. Zhang, 2020). Such conditions intensify the internalization of in-game information and its influence on players' value systems (Calleja, 2011).

*Theoretical and Practical Implications*

This study illustrates how GITT can be meaningfully extended into studying virtual ecologies. By integrating anthropocentric worldviews, exploitative gameplay behaviors, and immersive psychological states within the GITT framework, the research uncovers how virtual interactions simultaneously reflect and reshape human–nature relationships. From this dynamic, multi-level perspective, virtual environments emerge as both mirrors of preexisting worldviews and an infosphere for transforming environmental values. The findings also suggest potential behavioral spillover, where patterns of interaction developed in virtual contexts may carry over into real-life ecological behaviors. This resonates with theories of habit formation (Gardner & Lally, 2018) and cognitive rehearsal (Northam, 2000), implying that repeated virtual actions could gradually normalize specific attitudes toward nature. Finally, the study raises important considerations for virtual world design, highlighting how seemingly harmless gameplay mechanics can inadvertently shape players' environmental awareness and sensibilities.

Findings from this study suggest that video games such as ACNH simulate ecological systems that can serve as proxy ecosystems, enabling researchers to examine human–nature interactions in controlled, replicable environments. These digital ecosystems allow the exploration of behavioral responses to resource scarcity, punishment, and reward systems, mirroring real-world ecological dynamics. The study advances the discourse on anthropocentric mindsets by illustrating how virtual agency—the ability to exploit digital resources without tangible consequences—can either reinforce or challenge human-centered worldviews. Depending on the game's design and feedback mechanisms, immersive gameplay can desensitize players to ecological degradation or heighten their environmental awareness. This dual potential highlights new possibilities for designing virtual interventions that nurture the Nature Quotient (NQ) through meaningful, reflective play (Vuong & Nguyen, 2025).

Beyond entertainment, these insights have broad implications for nature-based digital interventions, gamified education, and digital health tools. This research underscores how digital behaviors can function as proxies for environmental values and demonstrates how game design can be leveraged as an educational and transformative tool. Developers can foster ecological awareness by integrating environmental feedback loops into gameplay—such as introducing visible consequences for overexploitation (e.g., ecosystem degradation or resource depletion) or rewarding sustainable behaviors like replanting, recycling, or habitat restoration. At the same time, designers must be mindful of how immersion and reward structures can inadvertently normalize exploitative mindsets or weaken pro-environmental values.

Policymakers can recognize virtual platforms as emerging informal learning spaces and develop frameworks to enhance NQ, and subsequently the eco-surplus culture (Nguyen, Ho, & La, 2025; Vuong, Nguyen, & La, 2025). At the same time, they should remain alert to how immersive elements may inadvertently reinforce exploitative mindsets or dilute pro-environmental attitudes. Establishing ethical design guidelines for nature-based games is therefore vital to prevent the normalization of extractive or harmful virtual behaviors. To address these issues comprehensively, several interconnected policy directions are proposed:

Governments should encourage developers to embed ecological narratives and ethical cues that cultivate empathy toward non-human life. Dynamic feedback systems should be promoted—rewarding conservation-oriented actions (e.g., replanting, recycling, habitat restoration) while discouraging exploitative behaviors. Ethical certification schemes for environmentally conscious games could further incentivize this approach. Moreover, Public awareness campaigns and educational programs should be developed to help users—especially young players—understand how immersive environments shape perceptions of nature and resource use. By illuminating the psychological gap between virtual and real ecosystems, such initiatives can nurture more reflective, responsible engagement with both.

Collaboration between game developers, educators, and environmental organizations can turn immersive platforms into powerful pedagogical tools. Virtual simulations can model the consequences of anthropocentric actions, fostering systems thinking, moral engagement, and ecological stewardship in learners. Public and private institutions should fund interdisciplinary research examining how virtual experiences shape ecological values and pro-environmental behaviors. Longitudinal and cohort studies can help identify how sustained interaction with virtual nature affects real-world sustainability practices, empathy, and ecological reasoning.

*Study Limitations*

This study is not without limitations. The dataset was derived from a cross-sectional design, which limits the ability to draw causal inferences and capture temporal dynamics. Consequently, it remains unclear whether virtual behaviors shape anthropocentric attitudes or whether preexisting attitudes influence virtual behaviors. Similarly, potential changes in players' mindsets over time or through repeated gameplay were not assessed, leaving longitudinal effects unexplored.

The data also rely on self-reported behaviors and perceptions from ACNH players, which may be influenced by recall bias or social desirability bias. Participants might have inaccurately estimated their in-game activities or underreported exploitative behaviors to align with perceived environmental norms. Additionally, the findings are context-specific to ACNH, limiting their generalizability to other virtual environments or game genres with different ecological dynamics. Interpretations of exploitative behaviors in this game may not apply to other games. Finally, the discussion of ACNH's punishment mechanisms remains theoretical reasoning and untested; without direct measures of players' awareness or responses to in-game ecological feedback, these mechanisms should be regarded as hypothetical pathways rather than confirmed causal processes.